\documentclass{book}

\usepackage{Wiley-AuthoringTemplate}
\usepackage{natbib}

\usepackage{ulem} 

\usepackage[capitalise,noabbrev]{cleveref}
\crefformat{equation}{Equation #2#1#3}
\begin{document}

\chapter[Human Behavioral Models Using Utility Theory and Prospect Theory]{Human Behavioral Models Using Utility Theory and Prospect Theory \protect}

\author*[1]{Anuradha M. Annaswamy}
\author[2]{Vineet Jagadeesan Nair}

\address[1]{\orgdiv{Department of Mechanical Engineering}, 
\orgname{Massachusetts Institute of Technology}, 
\postcode{02139}, \countrypart{MA}, 
     \city{Cambridge}, %\street{Massachusetts Avenue}, 
     \country{USA}}%

%\address[3]{\orgdiv{II Author Organization Division Name}, 
%\orgname{Organization Name}, 
%\postcode{Postal Code}, \countrypart{Part of the Country}, 
%     \city{City Name}, \street{Street Name}, \country{Country}}%

\address*{Corresponding Author: Anuradha Annaswamy; \email{aanna@mit.edu}}

\maketitle% This tag is required to print author and address in the output
%\tableofcontents
\begin{abstract}{Abstract}
Several examples of Cyber-physical human systems (CPHS) include real-time decisions from humans as a necessary building block for the successful performance of the overall system. Many of these decision-making problems necessitate an appropriate model of human behavior. Tools from Utility Theory have been used successfully in several problems in transportation for resource allocation and balance of supply and demand \citep{ben1985discrete}. More recently, Prospect Theory has been demonstrated as a useful tool in behavioral economics and cognitive psychology for deriving human behavioral models that characterize their subjective decision making in the presence of stochastic uncertainties and risks, as an alternative to conventional Utility Theory \citep{kahneman_prospect_2012}. These models will be described in this article. Theoretical implications of Prospect Theory are also discussed. Examples will be drawn from transportation use cases such as shared mobility to illustrate these models as well as the distinctions between Utility Theory and Prospect Theory. 

\end{abstract}

\keywords{Prospect theory, Behavioral modeling, Smart Cities, Transactive Control, Utility Theory}

\section{Introduction}
\begin{figure}[H]
  \centering
    \includegraphics[width=\linewidth]{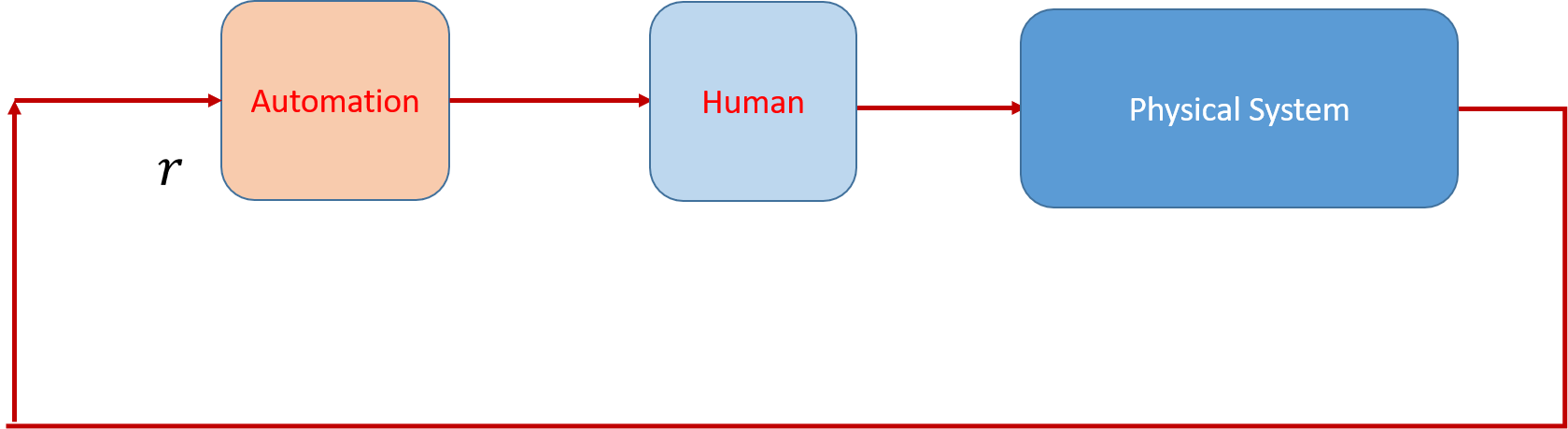}
  \caption{An Example of a CPHS with human in the loop.}
  \label{fig:CPHS}
\end{figure}

Analysis and synthesis of large-scale systems require the understanding of cyber physical human systems. Interactions between humans and cyber-components that interact with the physical system are varied and depend on a variety of factors and the goals of the large-scale systems. If the problem at hand concerns the behavior of the CPHS under emergency conditions, the interactions between humans and automation need to focus on a shared control architecture \citep{shared_Control_cphs} with appropriate granularity of task allocation and timeline. Typically the human in this context is an expert and when anomalies occur, either takes over control from automation or provides close supervision to the automation to ensure an overall safe CPHS. Under normal circumstances, the interactions may include other architectures. The role of the human is not necessarily that of an operator or an expert, but a user. The human may be a component in the loop, responding to outputs from the automation, and making decisions that serve in turn as inputs or reference signals $r$ to the physical system (see \cref{fig:CPHS}). 

Typical examples of such interactions have begun to occur both in power grids and transportation and can be grouped under the rubric of transactive control. 
%%all citations in these paragraphs till %%%% come from my paper in Proc. IEEE, “Transactive control in smart cities”; [*] corresponds to this Proc. IEEE paper. %%
The transactive control concept \citep{chassin2004modeling,bejestani2014hierarchical,annaswamy2015transactive} consists of a feedback loop resulting from incentives provided to consumers. Introduced in the context of smart grids, a typical transactive controller consists of an incentive signal sent to the consumer from the infrastructure and a feedback signal received from the consumer, and together the goal is to ensure that the underlying resources are optimally utilized. This introduces a feedback loop, where empowered consumers serve as actuators into an infrastructure, and transactive control represents a feedback control design that ensures that the goals of the infrastructure are realized, very similar to \cref{fig:CPHS}. The use of transactive control in smart grids can be traced to homeostatic control proposed in \citep{schweppe1978power} and \citep{schweppe1980homeostatic}, which suggested that demand-side assets can be engaged using economic signals. Transactive control in this context has come to denote market-based control mechanisms that incentivize responsive loads and engage them in providing services to the grid where and when of great need (as described in \citep{katipamula2006transactive,li2015market,hao2016transactive,somasundaram2014reference,hammerstrom2008pacific,widergren2014aep,melton2015pacific,kok2013powermatcher} and \citep{bernards2016meta}), and has the ability to close the loop by integrating customers via the right incentives to meet their local objectives such as lowering their electric bill as well as meeting global system objectives such as voltage and frequency regulation. Another example of transactive control is in the context of congestion control in transportation (\citep{phan2016model} and \citep{annaswamy2018transactive}) - where dynamic toll pricing, determined by the controller, is used as an incentive signal to the drivers, who then decide whether or not to enter a tolled segment, thereby regulating traffic density and possibly alleviating congestion. In both examples, it is clear that an understanding of the human behavior is important, and the modeling of the overall socio-technical system that includes the human behavior and their interaction with the physical system is an important first step in designing the feedback controller.  %%%%

This article focuses on two different tools that have been proposed for deriving behavioral models of human as a consumer. These tools include Utility Theory and Prospect Theory, and are described in the following sections. Prospect Theory is a framework introduced by Nobel prize-winning behavioral economists and psychologists that has been extensively shown to better represent decision making under uncertainty. It builds upon Utility Theory models by introducing additional nonlinear transformations on agents' objective utilities to model their irrational and subjective behaviors. Cumulative Prospect Theory (CPT) is an extension of Prospect Theory that also considers distortion of subjective probabilities through a weighting function. Examples are drawn from transportation to elucidate the impact of these tools, particularly in terms of predicting mode choice probabilities of passengers.
%How efficient and resilient perCPHS – interaction between humans and automation to improve performance of large-scale systems
%Several types of interactions. Can be traded. Can be supervisory. Can be shared. One of the interesting examples is to have a sequence of decision making as follows. Block diagram.
%Examples in power grid. Example in transportation. Transactive Control. In order to understand how to design the automation, socio-technical model. Interface with the system. Behavioral Model. The latter is the focus of this article. Utility Theory and Prospect Theory. Examples. 

\section{Utility Theory \label{sec:utility_theory}}
The starting point for Utility Theory \citep{ben1985discrete} is the assignment of a value to an outcome in the form of a utility function. If in general there are $N$ different possible choices, then $U_i$ is utility of outcome $i$, $i=1,\ldots, N$. The benefit of the utility function is that it provides a substrate for modeling a human’s decision when faced with  these choices. In particular, the probability that the human will select choice $\ell$ is determined using the utility function as:
\begin{equation}
    p^{\ell} = \frac{e^{U^{\ell}}}{\sum_{j=1}^N e^{U^{\ell}}} \; \ell \in \{1,\dots, N\}
    \label{eq:discrete_choice_model}
\end{equation}
Equation \ref{eq:discrete_choice_model} then serves as a simple behavioral model of the human which can be appropriately utilized in the overall problem of interest. For a simple problem with only two choices $A$ and $B$, the probability $p^A$ is given by
\begin{equation}p^A =\frac{1}{1+e^{-\Delta U}}\label{2}\end{equation} where $\Delta U=U^A-U^B$.
Modeling the behavior of the human in the problem, whether as a consumer, an operator, a user, or an expert, the Utility Theory based approach entails the characterization of all possible outcomes and the determination of the utility of each of the outcome $i$ as $U^i$. A variety of factors contributes to the utility and hence determining the behavioral model in \cref{eq:discrete_choice_model} is nontrivial. Quantitative aspects related to economy, qualitative aspects such as comfort, and hard to identify aspects such as strategic behavior, negative externalities, global and network based expectations, are all factors that may need to be simultaneously accounted for. Nevertheless, \cref{eq:discrete_choice_model} serves as a cornerstone for many problems where CPHS models have to be derived.

\subsection{An Example}
Suppose we consider a transportation example where a passenger intends to travel from point A to point B, and has two choices for transport: a shared ride service (SRS) and public transit. The utility function for this trip can be determined as:
\begin{align}
    u & = \mathbf{a}^\intercal \;\mathbf{t} + b\gamma + c \label{vineet1}
    \end{align}
where the components of $\mathbf{t} = [t_{walk}, t_{wait}, t_{ride}]^{\intercal}$ denote the walking, waiting, and riding times, respectively, $\gamma$ denotes the ride tariff, $\mathbf{a}=[a_{walk}, a_{wait}, a_{ride}]^{\intercal}$ are suitable weights, and $c$ denotes all other externalities that do not depend on either travel time or ride tariff. Both the weights $\mathbf{a}$ and tariff coefficient $b$ are assumed to be negative since these represent disutilities to the rider arising from either longer travel times or higher prices, while $c$ can be either positive or negative depending on the characteristics of the given travel option. All of the parameters ${\mathbf a}, b, c$ determine the behavior of a rider and could vary with time, the environment, or other factors. \cref{vineet1} indicates that the choice of the rider of the SRS over other options like public transit is determined by whether $u(SRS)>u(\hbox{public transit})$ for a given ride. 

A behavioral model of a driver as above was applied to the congestion control problem on a highway segment to determine a dynamic tolling price strategy \citep{ben1985discrete,phan2016model}. It was shown that with the alternative $u_0$ corresponding to travel on a no-toll road, the toll price can be determined using a nonlinear PI controller using a socio-technical model that was a cascaded system with the behavioral model as in \cref{2}-\ref{vineet1} and an accumulator model of the traffic flow. Using actual data from a highway segment in the US city of Minneapolis, it was shown that such a model-based dynamic toll price leads to a much more efficient congestion control \citep{annaswamy2018transactive}. More recently, this approach has been extended to a more complex highway section with multiple merges and splits, and applied to data obtained from a highway section near Lisbon, Portugal \citep{lombardi2021model}. Here too, a behavioral model of the driver similar to \cref{vineet1} was employed. The results obtained displayed a significant improvement compared to existing traffic flow conditions, with minimal changes to the toll price.

\section{Prospect Theory \label{sec:prospect_theory}}
The behavioral model based on Utility Theory have two deficiencies. The first is that the utility function is a embedded in a stochastic environment, causing the utility function model to be more complex than that considered in \cref{vineet1}. The second is that the model as considered in \cref{2} may not be adequate in capturing all aspects of decision making of a human. Strategic decision making, adjustments based upon the framing effect, loss aversion, and probability distortion are several key features related to subjective decision making of individuals when facing uncertainty. It is in this context that \textit{Prospect Theory} \citep{kahneman_prospect_2012} in general, and \textit{Cumulative Prospect Theory} (CPT) \citep{tversky_advances_1992}, in particular, provide an alternate tool that may be more appropriate. CPT builds upon Prospect Theory by using a probability weighting function to represent the agent's distortion of perceived probabilities of outcomes, and uses these probability weights to compute the subjective utilities from subjective values. We briefly describe this tool below. 

We first introduce a stochastic component into the problem and utilize the transportation example as the starting point. As travel times are subject to stochasticity, $u$ becomes a random process. For simplicity, suppose we assume that there are only two possible travel time outcomes, $\mathbf{\overline{t}}$ and $\mathbf{\underline{t}}$ ($\mathbf{\underline{t}} \leq \mathbf{\overline{t}}$) having corresponding utilities $\underline{u}$ and $\overline{u}$ ($\underline{u} \leq \overline{u}$), occurring with probabilities of $p \in [0,1]$ and $1-p$ respectively. It follows that the utility function for the SRS is given by
\begin{align}
\label{eq:utilities}
 \underline{u} & = \mathbf{a}_{sm}^\intercal \;\mathbf{\overline{t}} + b_{sm}\gamma_{sm} + c_{sm} \nonumber \\%= \underline{x} + b\gamma \\
 \overline{u} & = \mathbf{a}_{sm}^\intercal \;\mathbf{\underline{t}} + b_{sm}\gamma_{sm} + c_{sm} 
\end{align}
If these outcomes follow a Bernoulli distribution, its cumulative distribution function (CDF) is defined on the support $[\underline{u},\overline{u}]$:
\begin{equation}\label{cases}
    F_U(u) = \begin{cases} 0 &\mbox{if } u < \underline{u} \\
    p &\mbox{if } \underline{u} \leq u < \overline{u} \\
    1 &\mbox{if } u \geq \overline{u}. \ \end{cases}
\end{equation}

Suppose the alternative choice has a utility function $u_o$.
If $u_o \leq \underline{u}$, the customer would always choose the SRS since it offers strictly better outcomes and conversely if $u_o \geq \overline u$. For all other cases, the underlying model becomes a combination of \cref{2} and \cref{cases}. %Thus, the only cases considered are where $\underline{u} \leq u_o \leq \overline{u}$ are considered (note: $u_o$ can still be either a gain or loss) such that the consumer's choice (of accepting or rejecting the SRS ride offer) is non-trivial.

\begin{figure}[H]
  \centering
    \includegraphics[width=\linewidth]{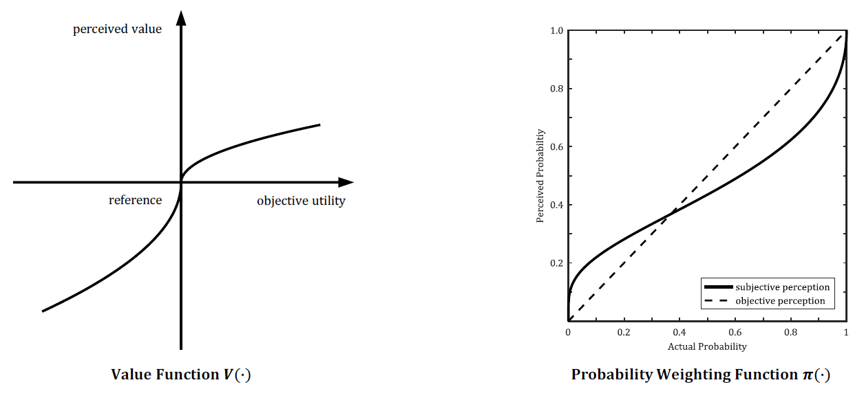}
  \caption{Illustrations of the CPT value function and probability weighting functions \citep{guan2019cumulative}.}
  \label{fig:CPTfuncs}
\end{figure}

We now address the second deficiency in Utility Theory. Conventional Utility Theory postulates that consumers choose among travel options based on their respective expected utilities \citep{fishburn1988nonlinear, von2007theory}. Alternatively, random utility models are another framework within Utility Theory that predict choice probabilities based on the utilities of different alternatives (computed using logit models) without accounting for risk, by assuming certain distributions for unobserved factors and error terms. However, both of these are inadequate when there is significant uncertainty involved. Prospect Theory (PT) is an alternative to Utility Theory that better describes subjective human decision making in the presence of uncertainty and risk \citep{tversky_advances_1992,kahneman_prospect_2012}, and Cumulative Prospect Theory (CPT) is a variant of PT that weighs different outcomes using distorted subjective probabilities as perceived by passengers. This is needed since individuals have been shown to consistently underestimate the likelihood of high probability outcomes while overestimating the likelihood of less likely events \citep{tversky_advances_1992}. To describe CPT, we introduce a value function $V(\cdot)$ and a probability distortion $\pi(\cdot)$ given by \citep{guan2019cumulative} and \citep{prelec_probability_1998}, with $\pi(0) = 0$ and $\pi(1) = 1$ by definition. These nonlinearities map the objective utilities ($u$) and probabilities ($p$) of each possible outcome to subjective values, as perceived by the passengers. Note here that the probability weighting function $\pi(\cdot)$ as described in \cref{eq:prob_distort} is unique to CPT. The graphs in \cref{fig:CPTfuncs} show examples of how the value and probability weighting functions may vary according to the objective utility $u$ and actual probability $p$, respectively.
\begin{align}
    V(u) & = \begin{cases} (u-R)^{\beta^+} &\mbox{if } u \geq R \\
    -\lambda(R-u)^{\beta^-} &\mbox{if } u < R
    \end{cases}
    \label{eq:value_func} \\
    \pi(p) & = e^{-(-ln(p))^\alpha} 
    \label{eq:prob_distort}
\end{align}

The CPT parameters here describe loss aversion ($\lambda$), diminishing sensitivity in gains ($\beta^+$) and losses ($\beta^\_$) and probability distortion ($\alpha$). The reference $R$ is the baseline against which users compare uncertain prospects. These can vary across individuals and also depending on the particular set of alternatives the customer is facing.

With the above distortions in the value function, the overall utility function for a given stochastic outcome gets modified. Noting that the utility function $U$ is a random variable, a Utility Theory based derivation is as follows.  If $U$ takes on discrete values $u_i \in \mathbb{R}, \forall i \; \in \{1, \, \, \dots, \, n\}$ and the outcomes are in ascending order, i.e., $u_1 < \cdots < u_n$, where $n \in {\mathbb{Z}}_{>0}$ is the number of possible outcomes, one can determine the objective utility $U^o$ as the expectation of $U$ according to Utility Theory as \citep{von2007theory}, i.e.,
\begin{equation}
	\label{EUT_calculate_discrete}
	U^o = \sum_{i=1}^n p_i u_i 
\end{equation} 
where $p_i \in (0, 1)$ is the probability of outcome $u_i$, and $\sum_{i=1}^n p_i=1$. 
In contrast, a Prospect Theory based derivation of utility function includes $V$ rather than $u_i$ and the probability distortion $\pi(.)$ applied to the probabilities $p_i$. Suppose we define the corresponding  utility function as $U^s_R$, which is the subjective utility perceived by the passenger according to Cumulative Prospect Theory, then
\begin{equation}
	\label{CPT_calculate_discrete}
	U^s_R = \sum_{i=1}^n w_i V(u_i)
\end{equation}  
where $R$ denotes the reference corresponding to the framing effect mentioned above, and $w_i$ denotes the weighting that represents the subjective perception of $p_i$. Suppose that $k$ out of the $n$ outcomes are losses, $0 \leq k \leq n, k \in \mathbb{Z}_{\geq 0}$, and the rest are non-losses, i.e., $u_i < R$ if $1 \leq i \leq k$ and $u_i \geq R$ if $k < i \leq n$. We can then derive the subjective probability weights assigned by the decision maker to each of these discrete outcomes from the cumulative distribution function of $U$ given by $F_U(u)$, as follows:
\begin{equation}
	\label{weights}
	w_i =
		\begin{cases}
			\pi\big[F_U(u_i)\big] - \pi\big[F_U(u_{i-1})\big], & \text{if} \; i \in [1, k] \; \text{(losses)} \\
			\pi\big[1-F_U(u_{i-1})\big] - \pi\big[1-F_U(u_i)\big], & \text{otherwise} \; \text{(non-losses)}
		\end{cases} 
\end{equation} 
where we have assumed  $F_U(u_0) = 0$ for ease of notation.

It is clear that in contrast to $U^o$, $U^s_R$ is centered on $R$, loss aversion is captured by choosing $\lambda > 1$, and diminishing sensitivity by choosing $0< {\beta}^+, {\beta}^- < 1$. The probability distortion is quantified by choosing $0 < \alpha < 1$. The extension from \cref{CPT_calculate_discrete} to the continuous case of $U^s_R$ is 
\begin{equation}
	\label{CPT_calculate}
	U^s_R = \int_{-\infty}^{R} V(u) \frac{d}{du}\Big\{\pi\big[F_U(u)\big]\Big\}du + \int_{R}^{\infty} V(u)\frac{d}{du}\Big\{-\pi\big[1-F_U(u)\big]\Big\}du
\end{equation}
 
With the above objective evaluation of a utility function as in \cref{EUT_calculate_discrete} and subjective evaluation as in \cref{CPT_calculate_discrete}, one can now determine the probability of acceptance of an outcome as follows. As has been shown in \cref{eq:discrete_choice_model}, the evaluation of the probability of acceptance of an outcome $\ell$ requires the utility of all alternate outcomes. Without loss of generality, suppose there are only two alternatives, with the objective and subjective utility of option $i \in \{1, 2\}$ given by $U_{io}$ and $U_{is_R}$ respectively. Then the objective probability of acceptance of choice $1$ using Utility Theory is given by
\begin{equation}
	\label{prob_a_binary_objective}
	p^o_1 = \frac{e^{U_{1o}}}{e^{U_{1o}}+e^{U_{2o}}}
\end{equation} 
while the subjective probability of acceptance of option $1$ using Prospect Theory is given by
\begin{equation}
	\label{prob_a_binary_subjective}
	p^s_1=\frac{e^{U_{1s_R}}}{e^{U_{1s_R}}+e^{U_{2s_R}}}
\end{equation}  
%where $A^s_R$ denotes the subjective utility of the alternative perceived by the passenger, which can be derived via (\ref{objective_utility}), (\ref{CPT_calculate_discrete}), and (\ref{V_function}). 

%\subsection{Determination of Parameters}
%\label{parameterization}

\subsection{An example: CPT modeling for SRS}

We illustrate the Prospect Theory model using the transportation example considered above, extended to the case where a passenger now has three choices $i \in \{1, 2, 3\}$, (i) public transit like buses or the subway, (ii) using a shared ride pooling service (SRS) and (iii) another which may be an exclusive ride hailing service (such as UberX). The SRS has greater uncertainty in pick up, drop off and travel times when compared to the UberX and transit alternatives, due to the possibility of more passengers being added en route. Thus, both UberX and transit can be treated as certain prospects when compared to the SRS option, which has uncertain outcomes. CPT can then be used to model the passenger's risk preferences to predict their decision making under such uncertainty. The discussions above show that a number of parameters related to the CPT framework have to be determined. These include $\alpha, {\beta}^+, {\beta}^-, \lambda$ defined in $V(\cdot)$ and $\pi(\cdot)$, which are in addition to the parameters $\mathbf{a}, b, c$ defined in (\cref{vineet1}) associated with the travel times $t_{\text{walk}}, t_{\text{wait}}, t_{\text{ride}}$ and tariff coefficients, for all three travel modes. In order to estimate these parameters, we designed and conducted a comprehensive survey eliciting travel and passenger risk preferences from $N = 955$ respondents in the greater Boston metropolitan area \citep{jagadeesan2021estimation}. Note that the constant terms in the utility function, $c_{UberX}$ and $c_{SRS}$ were measured relative to public transit as a baseline, i.e. $c_{transit} = 0$.

\begin{table}[h!]
\centering
\caption{\label{tab:mode_choice}Parameters describing the discrete choice logit models for SRS and UberX.}{
\begin{tabular}{@{}ccccc@{}}
\hline
\textbf{Parameter}      & \textbf{Mean} & \textbf{SE} & \textbf{SD} & \textbf{SE} \\ \hline
$a_{walk}$ [$min^{-1}$]            & -0.0586       & 0.0053      & 0.1412      & 0.0079      \\
$a_{wait}$ [$min^{-1}$]             & -0.0113       & 0.0182      & 0.1491      & 0.0356      \\
$a_{ride, \; transit}$ [$min^{-1}$]    & -0.0105  &   0.0013  &   0.0284   &  0.0017      \\
$a_{ride, \; UberX}$ [$min^{-1}$]    & -0.0086       & 0.0014      & 0.0058      & 0.0010      \\
$a_{ride, \; SRS}$  [$min^{-1}$]    & -0.0186       & 0.0013      & 0.0095      & 0.0007      \\
$b$ [$\$^{-1}$]                    & -0.0518       & 0.0050      & 0.0597      & 0.0042      \\
$c_{UberX}$             & -2.5926       & 0.1800      & 2.3034      & 0.1558      \\
$c_{SRS}$               & -2.2230       & 0.1497      & 1.8175      & 0.1530      \\ \hline
\end{tabular}}{}
\end{table}

\cref{tab:mode_choice} summarizes the mean values and standard deviations of the parameters that we estimated for the discrete mode choice model using maximum simulated likelihood estimation, along with their standard errors. We can also use the estimated travel time and price coefficients to determine the passengers' value of time (VOT) spent on different modes. The value of time is defined as the extra tariff that a person would be willing to pay or cost incurred to save an additional unit of time, i.e., it measures the willingness to pay (WTP) for extra time savings. In absolute terms, the VOT spent on mode $i$ can be calculated as the ratio between the marginal utilities of travel time and trip cost:
\begin{align}
    VOT_i & = \frac{\frac{\partial U_i}{\partial t}}{\frac{\partial U_i}{\partial \gamma}} = \frac{a_i}{b_i}
\end{align}

\begin{table}[h!]
\centering
\caption{\label{tab:value_of_time} Value of time spent on different modes, obtained from the random parameters logit model.}{
\begin{tabular}{@{}cc@{}}
\hline
\textbf{Trip leg or mode} & \textbf{VOT} (in \$/h) \\ \hline
Walking                    &   67.8702                                   \\
Waiting                    &  13.1480                                          \\
Transit ride               &  12.1703                                 \\
Exclusive ride hailing      &   9.9466                                      \\
Pooled ride sharing         &    21.5549                                       \\ \hline
\end{tabular}}{}
\end{table}

\begin{figure}[h!]
	\centering
	\includegraphics[width=0.9\textwidth]{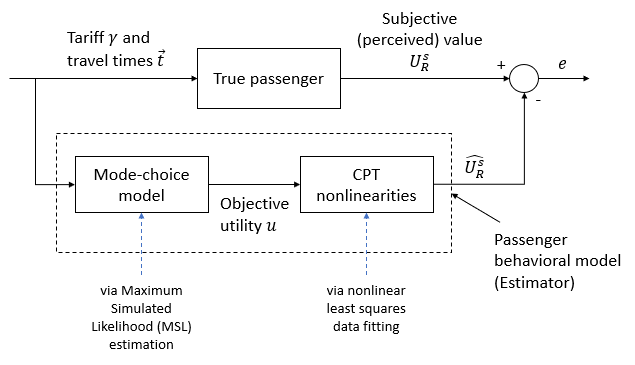}
	\caption{\label{fig:estimation} Estimation of mode choice models and CPT parameters from passenger survey data.}
\end{figure}

For the CPT model, estimated, we slightly modified the model to allow for different probability distortion effects in the gain and loss regimes. Thus, the modified weighting function is given by:
\begin{equation}
    \pi_{\pm}(p) = e^{-(-ln(p))^{\alpha_{\pm}} }
    \label{eq:orig_mod_weighting}
\end{equation}
The CPT risk parameters were estimated using the method of certainty equivalents \citep{rieger_estimating_2017,wang2019risk}, by presenting surveyed passengers with a series of chance scenarios asking them to choose a travel mode by comparing the uncertain or risky SRS versus the certain UberX and transit options. Nonlinear least squares curve fitting was then used to estimate the CPT parameter values. A schematic for the overall estimation process is shown in \cref{fig:estimation}. More details on the survey design and estimation procedures can be found in \citep{jagadeesan2021estimation}.

\begin{table}[h!]
\centering
\caption{\label{tab:chap4_cpt_lottery_final_stats} Summary of CPT parameter estimates for SRS travel preferences.}{
\begin{tabular}{@{}ccccccc@{}}
\hline
\textbf{}                   & $\alpha^+$ & $\alpha^-$ & $\beta^+$ & $\beta^-$ & $\lambda$ \\ \hline
\textbf{Mean}               &   0.4456    &   0.1315       &  0.2166 &    0.3550     &   20.0494        \\
\textbf{Median}             &  0.4124    &   0.1320       &  0.2188       &    0.3649    &   11.8715            \\
\textbf{SD}                 &   0.1828      &  0.0448      &   0.0985       &    0.1906     &  25.8554     \\ \hline
\end{tabular}}{}
\end{table}
Having estimated these parameters, we can compute the subjective or perceived value of the SRS outcomes, and the passenger's subjective probability of accepting the shared ride service offer using \cref{eq:value_func}-\ref{prob_a_binary_subjective}. Since the CPT model better accounts for passengers' irrational preferences when faced with uncertainty and risk, this subjective acceptance probability is more accurate than what would be predicted using conventional Utility Theory alone.

\subsubsection{Detection of CPT effects via lotteries}
In addition to estimating the numerical values parametrizing the subjective value and probability weighting functions, we also used the survey responses to detect the key CPT effects, which are:
\begin{itemize}
	\item \textit{Framing effect}: Individuals value prospects with respect to a reference point instead of an absolute value, and perceive gains and losses differently. 
	\item \textit{Diminishing sensitivity}: In both gain and loss regimes, sensitivity diminishes when the prospect gets farther from the reference. Therefore, the perceived value is concave in the gain regime and convex for losses, implying that people are risk-averse in gains and risk-seeking in the loss regime. 
	\item \textit{Probability distortion}: Individuals overweight small probability events and underweight large probability events.
	\item \textit{Loss aversion}: Individuals are affected much more by losses than gains. 
\end{itemize}
In order to estimate these effects, we considered simplified choice scenarios involving monetary lotteries. Survey respondents were also asked a series of hypothetical lottery questions after completing the SRS travel choice scenarios. We can then test for the existence of CPT-like behaviors based on their lottery responses, as described in \citep{rieger_estimating_2017}. More details on the survey and lottery scenarios can be found in \citep{jagadeesan2021estimation}.

\begin{table}[h!]
\centering
\caption{\label{tab:chap4_effects} Summary of key CPT effects observed from lotteries.}{
\begin{tabular}{@{}cc@{}}
\hline
\textbf{CPT effect tested}                    & \textbf{\% of valid responses} \\ \hline
Reflection effect (framing \& diminishing sensitivity)    & 95.03                          \\ \hline
\multicolumn{1}{c}{Probability overweighting between}                          \\ \hline
10\% and 60\% probability                     & 62.56 \%                       \\
60\% and 90\% probability                     & 40.51 \%                       \\
10\% and 90\% probability                     & 51.05 \%                       \\ 
Any probability weighting & 72.44 \%  \\ \hline
\multicolumn{1}{c}{Loss aversion}  \\ \hline
\multicolumn{1}{c}{Mean gain/loss ratio for mixed outcome lotteries}  & 3.7254 \\ 
Median gain/loss ratio for mixed outcome lotteries & 1.0250 \\
\hline
\end{tabular}}{}
\end{table}

From \cref{tab:chap4_effects}, we see that the valid responses clearly display CPT effects. The reflection or framing effect is shown by nearly all the valid respondents, indicating that our proposed value function is likely an accurate descriptor of how the passengers perceive their gains and losses. The probability weighting effect is not as dominant but it is still quite significant. We find that majority of them ($> 72 \%$) show at least some overweighting of probabilities and it is also most common in the lower probability ranges (between $10-60 \%$). This agrees with CPT theory since it postulates that people tend to overestimate the likelihood of rare events. The relatively large value of the mean gain/loss ratio ($> 1$) in the mixed lotteries indicates a significant degree of loss aversion among the surveyed passengers. However, the median value is quite close to 1 indicating that loss aversion may not be as prevalent for a sizeable portion of the passengers sampled in this study. These results from \cref{tab:chap4_cpt_lottery_final_stats} and \cref{tab:chap4_effects} demonstrate that we can use survey responses and data on passenger choices to (i) validate CPT behavioral effects and also (ii) estimate mathematical models and parameters to describe their risk preferences. Using real-time data from sources like ridesharing apps, these models can be continuously updated and improved in order to more accurately predict passenger choice and behavior.

\subsection{\label{sec:implications} Theoretical implications of CPT}

In the previous section, we determined mode choice and CPT behavioral models for passengers for a large population and over many different possible travel scenarios. In the following, we consider the case of a single trip (ride request and offer), in order to draw a few key insights about passenger risk preferences and travel behavior using computational experiments \citep{guan2019cumulative}. Here, the passenger makes a choice between SRS and an exclusive ride-hailing service like UberX.

% \begin{figure}[h!]
% 	\centering
% 	\includegraphics[width=0.9\textwidth]{table.png}
% 	\caption{\label{tab:scenario} Numerical values of parameters used in computational experiments in \cref{sec:implications}.}
% \end{figure}

% Table \ref{tab:scenario} summarizes the values of the parameters that we used in order to carry out the studies reported in this section. In particular, $\alpha$ was estimated from a recent survey study on passenger preferences under risk regarding transportation options conducted in Singapore involving $1,142$ participants with various demographics \citep{wang2018risk}, and $\beta^+, \beta^-, \lambda$ are from \citep{tversky_advances_1992}. In what follows, UberX is regarded as the alternative. The utility coefficients $[a_1, a_2, a_3, b, c]$ of both SRS and UberX were estimated from the same survey study in \citep{wang2018risk}. 
A dynamic routing problem of sixteen passengers using real request data from San Francisco was considered (see \citep{annaswamy2018transactive} for details), and the request from the $6^{\text{th}}$ passenger was used for the computational experiments in this section. The AltMin algorithm developed in \citep{guan2019dynamicrouting} was applied to derive the route and therefore the corresponding travel times of the SRS. The constraints on the possible delay were set to be at most 4 minutes of extra waiting and riding, respectively. For the same request, the travel times and price of the UberX option were retrieved from Uber\footnote{\url{https://www.uber.com/}}.

Using the travel times and prices, along with utility coefficients from \cref{tab:mode_choice}, the objective utility of UberX $A^o$ and $\underline{x}, \; \overline{x}$ of the SRS are calculated, using \cref{eq:utilities}. Note that $A^o, \underline{x}, \overline{x}$ are negative as they represent disutilities due to travel times and tariffs. Using this numerical setup, we explore the three implications via simulations: (i) fourfold pattern of risk attitudes, (ii) strong aversion of mixed prospects, and (iii) self reference. We use the following key properties of CPT based behavioral models for our analysis, more details and derivations of these can be found in \citep{guan2019cumulative}:

The first two properties are related to static and dynamic references, and are stated in Property \ref{monotonic_static} and \ref{monotonic_dynamic} below. These are helpful in determining the dynamic tariff $\gamma$ that allows $p^s_R$ to reach $p^*$, the desired probability of acceptance. Let $\bar{U} = \mathbb{E}_{f_U}(U)$ and $\bar{X} = {\mathbb{E}}_{f_X}(X)$, the third and fourth property stated in Property \ref{existence_lambda} and \ref{prob_a_for_mixed} are related to $U^s_{\bar{U}}$ and $p^s_{\bar{U}}$, respectively.
\begin{enumerate}
    \item \label{monotonic_static}
	Given any static reference point $R \in \mathbb{R}$, $p^s_R$ strictly decreases with $\gamma$.
	\item \label{monotonic_dynamic}
	Given any dynamic reference point in the form of $R = \tilde{x}+b\gamma, \tilde{x} \in \mathbb{R}$, $p^s_R$ strictly decreases with $\gamma$.
	\item 	\label{existence_lambda}
	Given any uncertain prospect, there exists a ${\lambda}^*$, such that $\forall \lambda > {\lambda}^*$, $U^s_{\bar{U}} < 0$.
	\item 	\label{prob_a_for_mixed}
	For any uncertain prospect, given that $\lambda$ is sufficiently large such that $U^s_{\bar{U}} < 0$, within the price range $\gamma \in [\underline{\gamma}, \overline{\gamma})$, where $\underline{\gamma}$ satisfies $\bar{X} + b \underline{\gamma} = A^o$, and $\overline{\gamma}$ satisfies ${\big[A^o - (\bar{X} + b \overline{\gamma})\big]} ^ {{\beta}^+} - U^s_{\bar{U}} = A^o - (\bar{X} + b \overline{\gamma})$, $p^s_{\bar{U}} < p^o$.
\end{enumerate}

\subsubsection{Implication I: Fourfold Pattern of Risk Attitudes}
\label{fourfoldpattern}

The fourfold pattern of risk attitudes is regarded as “the most distinctive implication of Prospect Theory” by Tversky and Kahneman \citep{tversky_advances_1992}, which states that when facing an uncertain prospect, the risk attitudes of individuals can be grouped into four categories: 
\begin{enumerate}
	\item Risk averse over high probability gains.
	\item Risk seeking over high probability losses.
	\item Risk seeking over low probability gains.
	\item Risk averse over low probability losses. 
\end{enumerate}
These risk attitudes are often used to justify subjective decision making of individuals for problems such as settlements of civil lawsuits, desperate treatments of terminal illnesses, playing lotteries, and getting insurance coverage. 

We now illustrate the fourfold pattern in the SRS context using the following scenario, which corresponds to the classic setup for the analysis of the fourfold pattern \citep{tversky_advances_1992}: Individuals decide between two options, a certain prospect and an uncertain prospect with two outcomes. The uncertain prospect is the SS, which we assume (i.e., UberX) obeys a truncated Poisson distribution with $K = 1$, i.e., the passenger is subject to at most one delay. Therefore, the two possible outcomes of the SRS are $(\underline{x} + b\gamma) $ and $(\overline{x} + b \gamma)$. The corresponding probabilities can be determined using:
\begin{equation}
	\label{define_Poisson}
	f_X^P(x)=
		\begin{cases}
			\frac{1}{Z^P}\frac{{{(\lambda}^P)}^k e^{-{\lambda}^P}}{k!}, \, & \text{if} \, x = \overline{x} - k \frac{\overline{x}-\underline{x}}{K}\\
			0, & \text{otherwise}
		\end{cases}
\end{equation}

\begin{equation}
	\label{fourfold}
	f_X^P(\underline{x}) = \frac{{\lambda}^P}{{\lambda}^P+1}, \quad f_X^P(\overline{x}) = \frac{1}{{\lambda}^P+1}
\end{equation}

\begin{figure}[h!]
	\centering
	\includegraphics[width=0.9\textwidth]{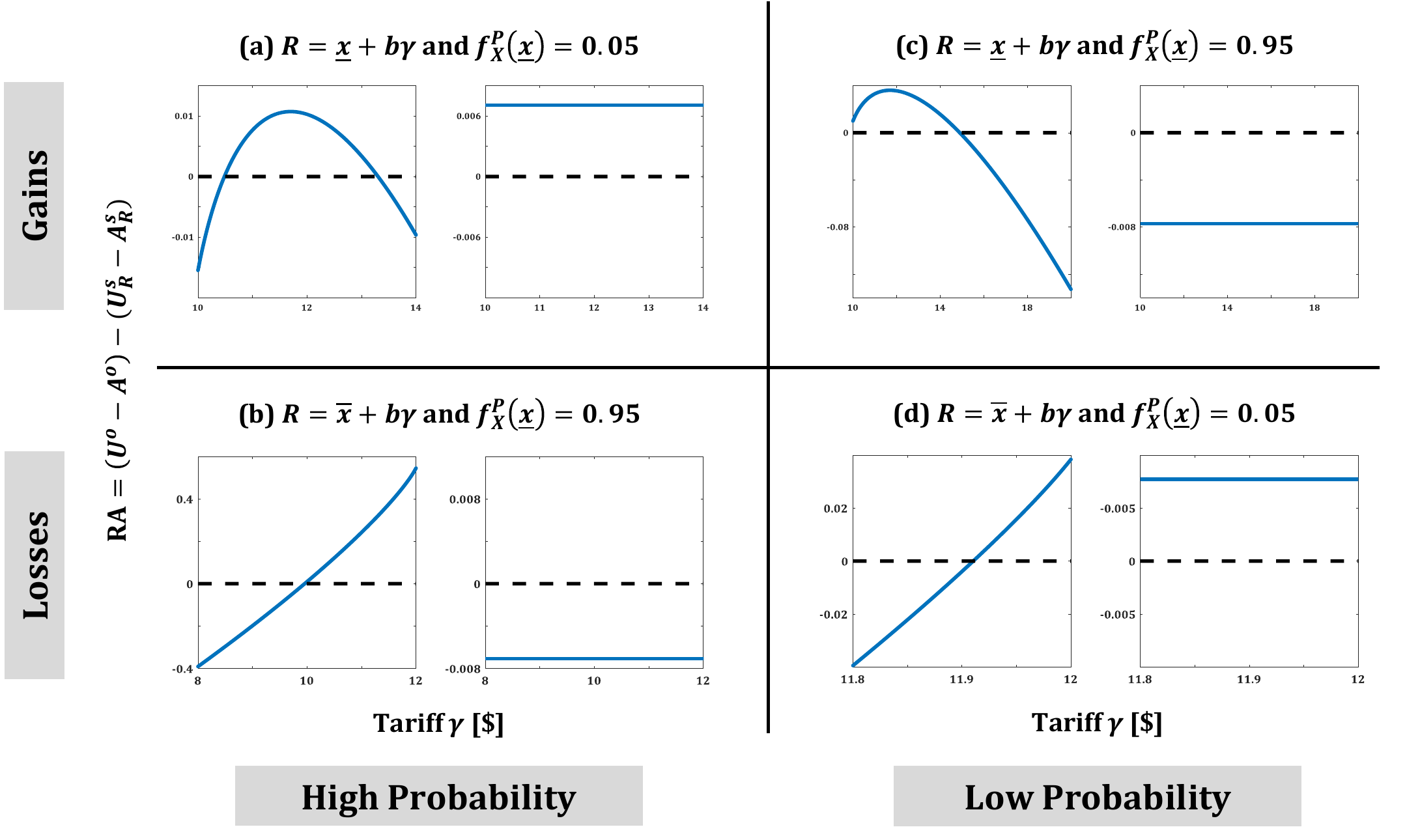}
	\caption{Illustration of the fourfold pattern of risk attitudes in the SRS context.}
	\label{fig:fourfold}
\end{figure}

The four scenarios above are realized through suitable choices of $R$ and ${\lambda}^P$ as follows. A dynamic reference point $R$ is chosen to be either $(\underline{x} + b\gamma)$ or $(\overline{x} + b\gamma)$, the SRS is a gain if $R = \underline{x} + b\gamma $ and a loss if $R = \overline{x} + b \gamma$. The SRS is considered high probability or low probability when the outcome that is not regarded as the reference can be realized with a probability of $p_{\text{NR}}$ or $(1-p_{\text{NR}})$ respectively, where $p_{\text{NR}}$ is close to 1. In the computational experiments presented in \cref{fig:fourfold}, $p_{\text{NR}}=0.95$. Moreover, the range of the tariff is chosen as follows 
\begin{equation}
	\label{price_in_fourfold}
	\begin{cases}
		\underline{x} + b\gamma < A^o & \text{if} \; R = \underline{x} + b\gamma \\
		\overline{x} + b\gamma > A^o & \text{if} \; R = \overline{x} + b\gamma
	\end{cases}	
\end{equation}
such that the objective utility of the certain prospect, $A^o$, lies in the same gain or loss regime as the SRS and therefore represents a reasonable alternative to the SRS.

With the uncertain and the certain prospect defined in the SRS context above, we illustrate the fourfold pattern in \cref{fig:fourfold} using four quadrants. According to the fourfold pattern (a)-(d) in \cref{fig:fourfold}, the diagonal quadrants should correspond to risk averse behavior while the off-diagonal ones are risk seeking. In each quadrant, we plot a metric defined as $\text{RA}=(U^o - A^o) - (U^s_R- A^s_R)$ with respect to the tariff $\gamma$. This metric captures the Relative Attractiveness that the uncertain prospect has over the certain prospect for rational individuals versus individuals modeled with CPT. This follows since according to \cref{prob_a_binary_objective} and \cref{prob_a_binary_subjective}, $\text{RA} >0 \Rightarrow p^o > p^s_R$. In \cref{fig:fourfold}, we note that $\text{RA}>0$ corresponds to all regions where the blue curve is above zero and indicates risk averse attitudes, as rational individuals have higher probability to accept the uncertain prospect than irrational ones. Similarly, $\text{RA} < 0$ corresponds to the blue line being below zero and denotes risk seeking attitudes. In each quadrant, two subplots are provided, where the subplot on the right corresponds to a specific set of parameters $\beta^+ = \beta^- = \lambda = 1$ which completely removes the role of $V(\cdot)$, while the subplot on the left corresponds to standard CPT parameters chosen in the range $0 < \beta^+, \beta^- < 1, \; \lambda > 1$, and therefore a general CPT model. And as explained before, each quadrant corresponds to a specific choice of $R$ and ${\lambda}^P$, which together determine if an outcome is a gain or loss, and whether with high or low probability. 

The most important observation from \cref{fig:fourfold} comes from the differences between the left and right subplots in each of the four quadrants. For example, from \cref{fig:fourfold}(a), all risk attitudes in the right subplot correspond to $\text{RA}>0$ and therefore risk averse, while those on the left are only risk averse for a certain price range. That is, the four fold pattern is violated in the left subplot. The same trend is exhibited in all four quadrants. This is because, the fourfold pattern is due to the interplay between $\pi(\cdot)$ and $V(\cdot)$ and is valid only when the magnitude of $\pi(\cdot)$ is sufficiently large relative to that of $V(\cdot)$, such that probability distortion dominates \citep{harbaugh2009fourfold}. This corresponds to the right subplots\footnote{The subplot on the right in each quadrant corresponds to the case where individuals are risk neutral in the gain or loss regimes separately, and loss neutral, then $\pi(\cdot)$ alone is sufficient to generate the fourfold pattern.} as well as the left subplots within certain price ranges. 

The implication that we obtain from the analysis of the fourfold pattern of risk attitudes is that the resulting four categories can suitably inform the dynamic pricing strategy in the SRS, through the left subplots. That is, it allows a quantification of two qualitative statements (1) the presence of risk seeking passengers gives flexibility in increasing tariffs, and (2) the presence of risk averse passengers requires additional constraints or bounds on reasonable tariffs.  

\subsubsection{Implication II: Strong Risk Aversion over Mixed Prospects}
\label{mixed}

The other implication of the CPT framework is strong risk aversion over mixed prospects. A mixed prospect is defined as an uncertain prospect whose portfolio of possible outcomes involves both gains and losses \citep{kahneman_prospect_2012,abdellaoui2008tractable}. Clearly, the uncertain prospect is always mixed when the reference point $R$ corresponds to its expectation (i.e., the expected value of its outcomes). The strong risk aversion of mixed prospects stems from loss aversion, as the impact of the loss component often dominates its gain counterpart. This implication will be illustrated below in the SRS context using two different interpretations. 

The first interpretation follows from Property \ref{existence_lambda}, which essentially states that when $R = \bar{U}$, the subjective utility is strictly negative for a sufficiently large $\lambda$. Therefore, with $R = \bar{U}$ and such a $\lambda$, the uncertain prospect is subjectively perceived as a strict loss. This has been verified numerically with ${\lambda} > 1$. Since the objective utility relative to the expectation is neutral, hence strong aversion is exhibited. 

The second interpretation follows from Property \ref{prob_a_for_mixed}, which essentially states that when Property \ref{existence_lambda} holds, within the tariff range $[\underline{{\lambda}}, \overline{{\lambda}})$, the uncertain prospect is less likely to be accepted by the CPT inclined passengers compared with the rational ones, as $p^s_{\bar{U}} < p^o$.

\begin{figure}[h!]
	\centering
	\includegraphics[width=0.9\textwidth]{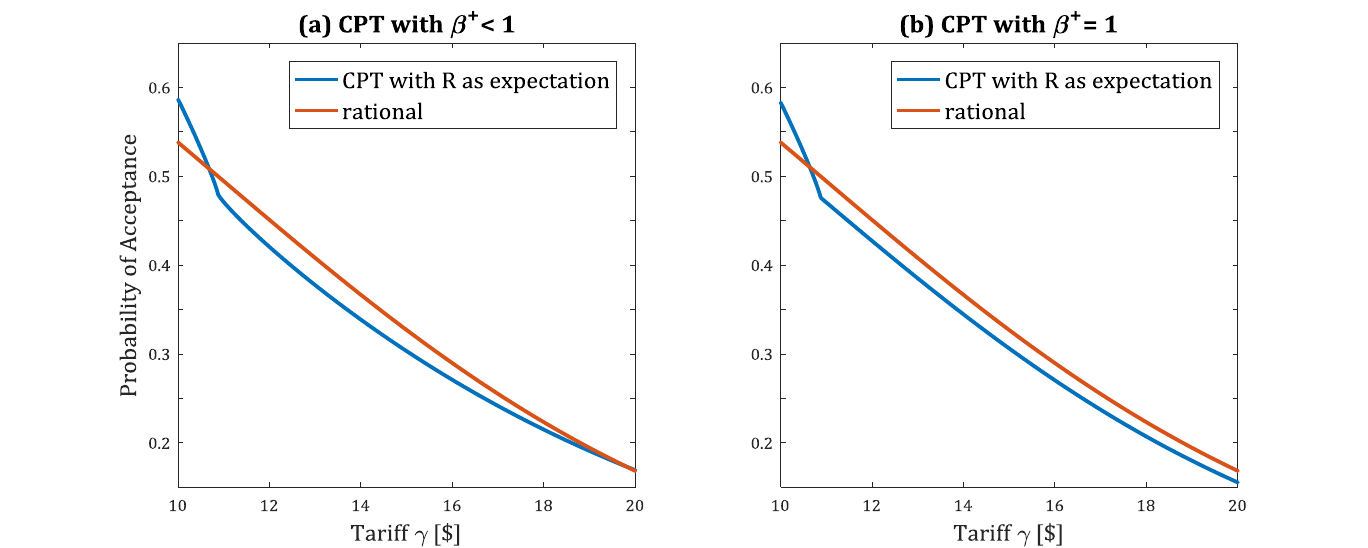}
	\caption{Comparison of $p^s_{\bar{U}}$ and $p^o$. For fair comparison, the tariff range of $\gamma \geq \frac{A^o - \bar{X}}{b}$ is plotted, where the alternative is non-loss.}
	\label{fig:mixed}
\end{figure}
Fig. \ref{fig:mixed} illustrates Property \ref{prob_a_for_mixed} with $f_X(x)$ obeying a Normal distribution, with the tariff range $\underline{\gamma}$ and $\overline{\gamma}$ approximated using the numerical setup. It is clear from the left subplot that within this price range, passengers exhibit strong risk aversion over the SRS, as the orange curve is strictly above the blue one. It is interesting to note that when ${\beta}^+ = 1$, which corresponds to the case when passengers are risk neutral in the gain regime, the maximum possible tariff $\overline{\gamma} \rightarrow \infty$ (see \cref{fig:mixed}(b)).  

The implication regarding strong risk aversion over mixed prospects is as follows: As the SRS has significant uncertainty, for passengers who regard the expected service quality as the reference, and when the alternative is relatively a non-loss prospect, strong risk aversion is exhibited. Hence the SRS is strictly less attractive to these passengers when compared to rational ones. Therefore, the dynamic tariffs may need to be suitably designed by the SRS server so as to compensate for these perceived losses. Rebates and subsidies may be a few typical examples. 

\subsubsection{Implication III: Self Reference}
\label{self}

\begin{figure}[h!]
	\centering
	\includegraphics[width=0.9\textwidth]{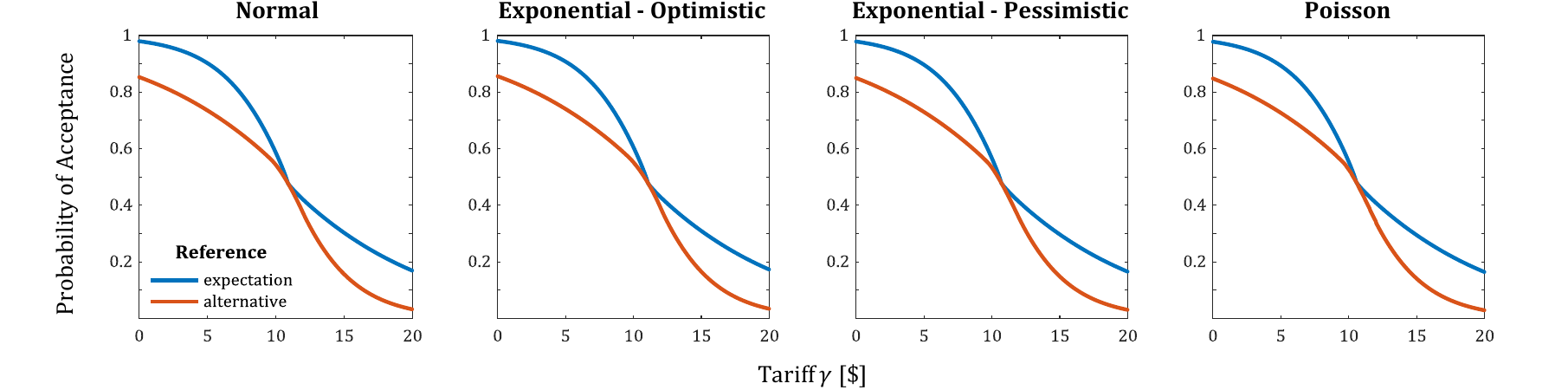}
	\caption{Comparison of $p^s_{\bar{U}}$ with $p^s_{A^o}$ using four different PDFs $f_X(x)$. In the truncated Poisson distribution, the parameters are set as ${\lambda}^P = 4$ and $K = 5$.}
	\label{fig:comparison}
\end{figure}

In this section, we compare $p^s_{\bar{U}}$ with $p^s_{A^o}$. Four different probability distributions are considered. In each case, how these two probabilities vary with the tariff $\gamma$ were evaluated. The results are shown in \cref{fig:comparison}.   

Fig. \ref{fig:comparison} illustrates that for all four distributions, $p^s_{\bar{U}} \geq p^s_{A^o}, \forall \; \gamma$, which implies that the SRS is always more attractive when the reference is the expectation of itself, rather than the alternative. $p^s_{\bar{U}} = p^s_{A^o}$ when $\gamma = \frac{A^o - \bar{X}}{b}$ therefore $\bar{U} = A^o$ hence the two reference points coincide. 

The following summarizes the third implication inferred from \cref{fig:comparison}: $\bar{U}$ is essentially the rational counterpart of the uncertain prospect. Therefore, it could be argued that, when deciding between two prospects, the chance to accept one prospect is always higher if this prospect itself is regarded as the reference, compared with the case where the alternative is considered as the reference. This is due to loss aversion, i.e., $\lambda > 1$, and can be explained thus: When one prospect is regarded as the reference, by definition, it would never be perceived as a loss and therefore not experience the magnified perception out of losses, whereas the alternative may be subject to being regarded as a loss and therefore can experience this skewed perception. In contrast, if the alternative is chosen as the reference, the roles are reversed\footnote{Other effects of CPT due to $\alpha, {\beta}^+, {\beta}^- < 1$ may result in complicated nonlinearities which might alleviate loss aversion. Therefore, this statement is valid when $\lambda$ is sufficiently large, such that loss aversion dominates.}. Moreover, the statement is in fact intuitive as those passengers who regard the expectation as the reference have in some sense already subscribed to the SRS, hence are naturally inclined to exhibit a higher probability of acceptance and therefore have higher willingness to pay. This partially explains the reason why converting customers from competitors is typically more difficult than maintaining the current customer base. The last observation from \cref{fig:comparison} is the invariance of the comparison with the underlying probability distributions, which implies that the above implications on self reference are fairly general. 
% \subsection{Dynamic Tariff Design}
% \label{design}
% With the above analytical properties of the CPT based passenger behavioral model, we propose the following algorithm for determining the dynamic tariff. The overall goal is for the actual probability of acceptance $p^s_R$ to reach the desired value $p^*$. We note from \cref{prob_a_binary_objective} that $p^s_R$ is a function of $U^s_R$ and $A^s_R$, which in turn is a function of $U$ following \cref{CPT_calculate_discrete}-\ref{CPT_calculate}. Finally, \cref{vineet1} shows that $U$ is a function of $\gamma$. By combining these equations, we can derive the relationship between $p^s_R$ and $\gamma$ as $p^s_R = f(\gamma)$. According to Property \cref{monotonic_static} and \cref{monotonic_dynamic}, $f(\cdot)$ is strictly monotonic. This in turn implies that the desired dynamic tariff that leads to $p^*$ is given by $\gamma = f^{-1}(p^*)$.
\section*{Summary and conclusions}
Several examples of CPHS include real-time decisions from humans as a necessary building block for the successful performance of the overall system. Many of these problems require behavioral models of humans that lead to these decisions. In this article, we describe two different tools that may be suitable for determining these behavioral models, which include Utility Theory and Prospect Theory. Tools from Utility Theory have been used successfully in several problems in transportation such as resource allocation and balance of supply and demand. This theory is described in \cref{sec:utility_theory} and illustrated using a transportation example that consists of a shared mobility problem where human riders are presented with the choice of different travel modes. We then show how these models can be used to address and mitigate traffic congestion. Cumulative Prospect Theory, an extension of Prospect Theory, is a modeling tool widely used in behavioral economics and cognitive psychology that captures subjective decision making of individuals under risk or uncertainty. CPT is described in \cref{sec:prospect_theory}, with the same transportation example used to illustrate its application potential. Results from a survey conducted with about 1000 respondents are used to derive a CPT model and estimate its parameters.  Lottery questions were also included in the survey to illustrate CPT effects, the results of which are described in this section as well. Finally, a few theoretical implications of CPT are presented that provide an overall quantitative structure to the qualitative behavior of humans in overall decision making scenarios. 

Human-in-the-loop behavioral models can be applied to several other applications beyond ridesharing, both within the transportation and mobility sector, and other domains that involve humans interacting with cyber-physical systems. One such example includes demand response in power grids. Future research involves the development of more accurate models of human operators for which more quantitative data is required with human decision makers as integral components of the overall infrastructure. For instance, these could include more sophisticated utility functions for different modes that also take into account factors other than price and travel time which could affect passengers' choices. In addition, more realistic data sets on ridesharing, mode choice and dynamic pricing could be obtained either by conducting larger, representative surveys or through pilot studies and field trials performed in conjunction with ridesharing companies or transit authorities. 

\section*{Acknowledgments}
This work was supported by the Ford-MIT Alliance. 

\backmatter

\normalem
\bibliography{references}

\end{document}